\documentclass[symmetry,article,accept,pdftex,moreauthors]{Definitions/mdpi} 
\firstpage{1} 
\pubvolume{1}
\issuenum{1}
\articlenumber{0}
\pubyear{2025}
\copyrightyear{2025}
\externaleditor{Firstname Lastname} % More than 1 editor, please add `` and '' before the last editor name
\datereceived{27 February 2025 } 
\daterevised{15 March 2025 } % Comment out if no revised date
\dateaccepted{16 March 2025 } 
\datepublished{ } 
\hreflink{https://doi.org/} % If needed use \linebreak
\usepackage[]{graphicx}
\usepackage[]{color}
\usepackage{alltt}
\usepackage{amsmath}
\usepackage{amssymb}
\usepackage{verbatim}
\usepackage{authblk}
\usepackage[nottoc]{tocbibind}
\usepackage{url}
\usepackage{amsfonts}
\usepackage{mdframed}
\usepackage[utf8]{inputenc}
\usepackage{ragged2e}

\usepackage{varioref}       

\newenvironment{eqaed}
    {\begin{equation}
    \begin{aligned}
    }
    { 
    \end{aligned}
    \end{equation}
    \ignorespacesafterend
    }

\usepackage{pgf}

\Title{Center Symmetry Breaking in Calabi--Yau Compactifications} %Title altered

\TitleCitation{Center Symmetry Breaking in Calabi--Yau Compactifications}

\Author{{Ivano Basile}
 $^{1,2,}$*\orcidA{} and Pouya Golmohammadi $^{2}$}

\AuthorNames{Ivano Basile and Pouya Golmohammadi}

\isAPAStyle{%
       \AuthorCitation{Basile, I., Golmohammadi, P.}
         }{%
        \isChicagoStyle{%
        \AuthorCitation{Lastname, Firstname, Firstname Lastname, and Firstname Lastname.}
        }{
        \AuthorCitation{Basile, I.; Golmohammadi, P.}
        }
}

\address{%
$^{1}$ \quad {Max-Planck-Institut f\"{u}r Physik (Werner-Heisenberg-Institut),} 
 Boltzmannstr. 8, 85748 Garching, Germany\\
$^{2}$ \quad Arnold-Sommerfeld Center for Theoretical Physics, Ludwig Maximilian University of Munich, Theresienstraße 37, 80333 Munchen, Germany; {pouya.golmohammadi@physik.uni-muenchen.de}
}

\corres{Correspondence: ibasile@mpp.mpg.de; Tel.: +49{-}
89{-}32354-267}

\abstract{It is widely believed that global symmetries must be broken in Quantum Gravity. This includes higher-form symmetries, which are commonplace in supergravity coupled to vector multiplets. Recently, a quantitative criterion for the breaking of (higher-form) symmetries in effective field theories of gravity has been proposed. We studied this criterion in the context of center one-form symmetries broken by BPS states in Calabi--Yau compactifications of type IIA string theory and M-theory. In a simple toy model, we evaluated the parameters quantifying the extent of symmetry breaking for large and small values of the moduli, comparing the scales of significant breaking with other relevant physical scales.}

\keyword{swampland program; string theory; generalized symmetries; supergravity; Calabi--Yau compactifications} 

\begin{document}

%%%%%%%%%%%%%%%%%%%%%%%%%%%%%%%%%%%%%%%%%%
 %% Remove this when starting to work on the template.
%\endnote{This is an endnote.} % To use endnotes, please un-comment \printendnotes below (before References). Only journal Laws uses \footnote.

% The order of the section titles is different for some journals. Please refer to the "Instructions for Authors” on the journal homepage.

\section{Introduction}

Symmetries have always been at the center of all aspects of physics. They provide physical insight into the behavior of complex systems, simplify computations in several instances, and~are intimately linked with conservation laws. The~connection between symmetries and gauge redundancies paved the way to a robust and parsimonious approach to model building in particle physics as well as condensed-matter physics. In~the past decade, especially in the context of (quantum) field theory, novel formal developments in the theory of symmetries~\cite{Gaiotto:2014kfa} (see~\cite{Schafer-Nameki:2023jdn, Brennan:2023mmt, Bhardwaj:2023kri, Shao:2023gho} for reviews) have spurred a series of breakthroughs~\cite{Choi:2022jqy, Putrov:2023jqi, Cordova:2022rer, Cordova:2024ypu, Bhardwaj:2023fca, Bhardwaj:2024qrf}, even leading to novel connections with experiments in high-energy and condensed-matter physics alike~\cite{Choi:2023pdp, Dierigl:2024cxm, Warman:2024lir}.

The simplest generalization of the ordinary notion of symmetry is the center one-form symmetry
exhibited by gauge theories. Much like local operators can be charged under an ordinary symmetry, such that their charge can be measured via Ward identities by linking their support with a codimension-one hypersurface (say a sphere $S^{d-1}$ in $d$ dimensions) supporting the symmetry generator, Wilson and 't Hooft lines carry a charge that can be measured by codimension-two linking hypersurfaces. The~corresponding symmetry group is the center $Z(G)$ of the gauge group. In~particular, Maxwell theory in four dimensions has a $U(1)_\text{el} \times U(1)_\text{m}$ \emph{{one-form}} symmetry, meaning that the conserved currents $F$ and $\star F$ are two-form rather than the usual one-form Noether currents. In~this framework, familiar statements such as gauge redundancy and the masslessness of photons, confinement, and Coulomb and Higgs phases can be recast in the language of symmetries, phase transitions, and order parameters~\cite{Gaiotto:2014kfa}.

Notwithstanding this essential role of symmetries in theoretical physics, in~the context of Quantum Gravity, the above considerations are reframed in a different perspective. Indeed, there are several motivations supporting the idea that Quantum Gravity is devoid of any and all symmetries~\cite{Misner:1957mt, Banks:2010zn, Harlow:2018tng, Harlow:2018jwu, McNamara:2019rup, McNamara:2020uza, McNamara:2021cuo, McNamaraThesis, Yonekura:2020ino, Bah:2022uyz, Heckman:2024oot, Bah:2024ucp}. These arguments range from considerations on semiclassical black-hole physics, topology fluctuations (via gravitational solitons and wormholes), holography, and string theory. In~the latter setting in particular, a~detailed understanding of how symmetries are broken has been developed, including the especially robust topological symmetries~\cite{McNamara:2019rup, Heidenreich:2020pkc}. Although~these kinematic results provide a satisfactorily coherent picture of the story, a~deeper understanding of these core aspects of Quantum Gravity must include dynamics. In~particular, if~symmetries ought to be broken, are there any bounds on the scale at which breaking effects would manifest and become significant? Must such a scale lie below the Planck scale $M_\text{Pl}$---or perhaps some other cutoff---or could symmetry-breaking rather occur in the fully non-perturbative regime in which spacetime itself is likely ill-defined? As a first step to begin approaching these questions, it is paramount to \emph{{quantify the extent to which symmetries are broken}}. Intuition from scattering amplitudes in gravity~\cite{Bah:2022uyz} suggests that a lower bound for symmetry-breaking effects would come from non-perturbatively suppressed contributions such as gravitational instantons. One would be thus tempted to take expressions such as $e^{-\frac{M_\text{Pl}^2}{E^2}}$ as some sort of minimal measure of symmetry breaking in Quantum Gravity. These heuristics already show that any such notion should depend on some typical energy scale at which the symmetry-breaking process is probed. From~an effective field theory point of view, another approach toward defining the extent to which a symmetry is broken begins from Ward identities for (higher-form) symmetries. When recast in an integrated form ({{this}
 approach also has the advantage of encompassing symmetries without a current}), symmetry operators $U(S^{d-p-1})$ supported on spheres linking the support of charged operators $V(\mathcal{C}^p)$ act by multiplication by a constant phase. In~this fashion, one can rephrase the conservation law as the statement that the symmetry operator $U$ be \emph{{topological}}, that is, its support can be deformed without changing observables, insofar as no other operator insertions are crossed. This contemporary language to frame symmetries then points to a suggestive way to quantify symmetry breaking: violating the topological property of symmetry operators is tantamout to saying that their action on charged operators depends on the geometry of their support. In~other words, the~phase given by linking $\mathcal{C}^p$ by a sphere $S^{d-p-1}$ depends (at least) on the radius of the sphere. Therefore, a~natural quantity that encodes the breaking of the corresponding Ward identity is just the (logarithmic) derivative of the effective charge that appears in the phase~\cite{Cordova:2022rer}. This strategy leads to a scale-dependent notion, since evaluating this derivative at a radius $r$ introduces a natural energy scale $\Lambda = r^{-1}$. Moreover, in~several examples, this notion reproduces the expected behavior when Ward identities are probed in this fashion at scales much smaller than the relevant scales of symmetry-breaking ingredients~\cite{Cordova:2022rer}.

In the framework developed in~\cite{Cordova:2022rer}, adapting the kinematic arguments to the effect that global symmetries are absent in Quantum Gravity leads to a more refined dynamical statement. Namely, as~we shall review in more detail in the following,~\cite{Cordova:2022rer} proposed that, in~effective field theories coupled to gravity, global symmetries ought to be \emph{{badly broken}} by the time one probes them at the scale where Quantum Gravity effects become significant. The~main goal of this paper is to study this proposal in a concrete top-down context provided by string theory. A~particularly relevant class for Quantum Gravity in this context is the breaking of center one-form symmetries of Abelian gauge sectors via charged massive matter. Partly, this is due to the fact that such settings commonly arise in the string landscape. More importantly, it is also because of its rich connections with the swampland program~\cite{Vafa:2005ui, Brennan:2017rbf, Palti:2019pca, vanBeest:2021lhn, Grana:2021zvf, Agmon:2022thq}. This is especially evident for the weak gravity conjecture~\cite{Arkani-Hamed:2006emk, Harlow:2022ich}; an~averaged version thereof was derived, suggesting that the corresponding one-form symmetry be badly broken at the appropriate scale~\cite{Cordova:2022rer}.

In this paper, we further develop this framework of the quantitative symmetry breaking of center one-form symmetries, extending its scope to general dimensions and massive species. We then study the proposal of~\cite{Cordova:2022rer} that such symmetries be badly broken in effective supergravity theories with eight supercharges in four and five dimensions, respectively obtained by compactifying type IIA string theory and M-theory on Calabi--Yau manifolds. While this is not the only option to realize these settings from string theory, it provides a convenient physical picture of the breaking mechanism at~play.

The contents of this paper are structured as follows. In~\cref{sec:approx_symms}, we introduce the notion of badly broken global symmetries and quantify the extent of symmetry breaking via the parameter defined in~\cite{Cordova:2022rer}. In~\cref{sec:bb_dd}, we describe detailed, one-loop computations of this parameter for charged scalars and spinors in diverse dimensions. In~\cref{sec:bb_BPS}, we study the behavior of this parameter and the proposal of~\cite{Cordova:2022rer} in four- and five-dimensional supergravity theories obtained compactifying type IIA string theory and M-theory on a Calabi--Yau threefold. In~order to carry out computations in closed form, we make several simplifying assumptions along the way, and~we discuss their limitations and workarounds for future work. Finally, in~\cref{sec:conclusions}, we provide some closing remarks and an outlook with several directions to pursue. In \cref{app:A} and \cref{app:B} we provide some background material and conventions on the relevant aspects of supergravity employed in the main~text.

%%%%%%%%%%%%%%%%%%%%%%%%%%%%%%%%%%%%%%%%%%
\section{Approximate Global Symmetries and Bad-Breaking~Parameter}\label{sec:approx_symms}
As stated earlier, one of the most well-established Swampland conjectures is the absence of global symmetries in Quantum Gravity. Although~these statements are conceptually far-reaching, they do not usually lead to concrete conclusions about specific phenomenological issues, such as the shift symmetry of an axion field, inflation in the early universe, the~lepton number in the standard model, and so on. The~basic reason is that the absence of global symmetry is a kinematic statement that does not significantly constrain the scale at which symmetries would break. One of the primary goals of this study was to develop a framework that enables us to quantify the notion of symmetry breaking. This was introduced in~\cite{Cordova:2022rer}, and,~in this work, we take the viewpoint discussed~therein.

\textls[-15]{To accomplish this objective and further clarify the concept of symmetry breaking in Quantum Gravity, we revisit the original remnant argument for the breaking of global symmetries, except~that allowing that the corresponding conservation law no longer holds at and above some UV scale $\Lambda_\text{bb}$. From~this point on, we will refer to these types of symmetries as \emph{{approximate}} global symmetries and~$\Lambda_\text{bb}$ as the \emph{{bad-breaking scale}}. One can immediately conclude that if the breaking scale is larger than the Planck scale or the breaking does not transpire below the Planck scale, the~Bekenstein--Hawking entropy bound will be violated. Consequently, it would be appear natural to presume that symmetry-violating effects begin to manifest significantly at or below the Planck scale ({{see}~\cite{Etheredge:2025rkn} for a recent study of more subtle settings, where apparent counterexamples can be understood as emergent gauge~redundancies}).}%
\begin{eqaed}\label{eq:bb_condition}
    \Lambda_\text{bb} \lesssim M_\text{Pl}.
\end{eqaed}
{Equivalently,}
 in~the scenario in which black-hole evaporation violates the symmetry, breaking effects start to kick in before the Schwarzschild radius $r_\text{S}$ of the black hole reaches a Planckian size, $r_\text{S}\sim l_\text{Pl}$. In~other words, these effects have non-gravitational origins within the domain of low-energy effective field theory. More generally, the~scale at which Quantum Gravity effects irredeemably break the effective field theory is the species scale $\Lambda_\text{sp}$ \cite{Dvali:2001gx, Veneziano:2001ah, Dvali:2007hz, Dvali:2007wp, Dvali:2009ks, Dvali:2010vm, Dvali:2012uq, vandeHeisteeg:2022btw, Cribiori:2022nke, Castellano:2022bvr, Blumenhagen:2023yws, Cribiori:2023sch, vandeHeisteeg:2023dlw, vandeHeisteeg:2023ubh, Castellano:2023aum, Castellano:2024bna, ValeixoBento:2025iqu, Calderon-Infante:2025ldq}, which raises the question of whether any consistent effective field theory coupled to gravity ought to satisfy $\Lambda_\text{bb} \lesssim \Lambda_\text{sp}$ instead of \cref{eq:bb_condition}. We shall return to this point in \cref{sec:bb_BPS}.

This picture can be straightforwardly extended to comprise higher-form symmetries, which is expected since ordinary global symmetries are regarded as $0$-form symmetries in modern terminology. Remarkably, one of the upshots of the analysis of~\cite{Cordova:2022rer} is that imposing such a constraint for higher-form symmetries will give rise to an averaged version of the weak gravity conjecture, although~we will not cover those arguments~here.

\subsection*{Bad-Breaking~Parameter}

In order to properly quantify the extent to which a symmetry is broken, the~starting point is to ask how the associated charge $q$ behaves in the presence of symmetry-breaking effects and across different regimes. In~the absence of these effects, the~Ward identity for (higher-form) symmetries can be expressed in terms of operatorial linking~\cite{Costa:2024wks}:
\begin{equation}
    U_{\alpha}(S^{(d-p-1)})V(\mathcal{C}^{(p)})=\exp(i\alpha q)V(\mathcal{C}^{(p)}),
\end{equation}
where the symmetry-generating topological operator $U_{\alpha}(S^{(d-p-1)})$ acts on the charged operator $V(\mathcal{C}^{(p)})$, which is supported on a $p$-dimensional manifold $\mathcal{C}^{(p)}$ linking the~sphere.

In the presence of charge-violating effects, the~symmetry operator loses its topological nature and the Ward identity no longer holds. Consequently, we anticipate that the charge will become dependent on the radius of the sphere $S^{(d-p-1)}$. With~this in mind, one can extract this effective charge according to~\cite{Cordova:2022rer}
\begin{equation}
    \lim_{\alpha \to 0}\frac{1}{i\alpha}\log \left(\frac{\langle V^{\dagger}(\mathcal{C}^{(p)})(\infty)U_{\alpha}\left(S_{r}^{d-p-1}\right)V(\mathcal{C}^{(p)})(0)\rangle }{\langle V^{\dagger}(\mathcal{C}^{(p)})(\infty)V(\mathcal{C}^{(p)})(0)\rangle}\right)= q(r),
\end{equation}
where, to avoid the denominator vanishing, we choose the locations (denoted schematically by $0$ and $\infty$) of charged operator insertions accordingly.
In the IR regime $\Lambda_\text{bb} r \gg 1$, we expect the topological feature of the symmetry operator to be restored. Put differently, as~we probe the theory at long distances, the~dependence of charge $q(r)$ on the radius should diminish, approaching the value $q_\infty$ attained in the unbroken theory. These observations lead us to introduce a notion that will henceforth be referred to as the ``{bad-breaking parameter}
'':
\begin{itemize}
    \item \textbf{{Bad-Breaking Parameter:}
}\textit{ {At}
 an energy scale $\Lambda = r^{-1}$, the~bad-breaking parameter denoted by $\delta_{\Lambda}$ quantifies the symmetry-breaking process in the effective field theory. Denoting the effective charge by $q(r)$ and its IR value $q_{\infty}$, we define}
\end{itemize}
\begin{equation}
    \delta_{\Lambda}\equiv \frac{r}{q_{\infty}}\frac{d}{dr}q(r).
\end{equation}
We say that the higher-form symmetry is badly broken at the scale $\Lambda$ if $\delta_{\Lambda} = O(1)$. Conversely, the~bad-breaking scale $\Lambda_\text{bb}$ is defined parametrically as the scale at which $\delta_{\Lambda_\text{bb}} = O(1)$.

To analyze the behavior of the bad-breaking parameter for symmetries that are explicitly broken at the Lagrangian level, we need to investigate the modification of the current at short distances. To~illustrate this scenario, consider a local operator $\mathcal{O}$ with scaling dimension $\Delta_{\mathcal{O}}$ that breaks the conservation of a current $J$ according to
\begin{equation}
    \delta J=\frac{\mathcal{O}}{\Lambda_\text{bb}^{\Delta_{\mathcal{O}}-d+1}} \,.
\end{equation}
Breaking effects are thus insignificant at low energies relative to the breaking scale $\Lambda_\text{bb}$. On~the other hand, at~short distances, the~effective charge is modified according to~\cite{Cordova:2022rer}
\begin{equation}
    \delta q(r)\sim \frac{1}{(r\Lambda_\text{bb})^{\Delta_{\mathcal{O}}-d+1}}\, ,
\end{equation}
confirming the expectation that the bad-breaking scale is in fact $\Lambda_\text{bb}$. More generally, the~relationship between extrapolated current and the symmetry-breaking scale can be more intricate, depending on the case at~hand.

In this paper, we shall focus on another mechanism that generates scale-dependence in the effective charge, namely, screening by vacuum polarization. This leads to terms in $q(r)$ that display non-analytic dependence on $r$. This is the relevant case for the breaking of electric and magnetic one-form symmetries studied in~\cite{Cordova:2022rer}. As~outlined in the introduction, this setting affords a particularly convenient top-down realization in string theory via BPS states in Calabi--Yau compactifications. On~general grounds, the~effective charge for the one-form symmetry acting on Wilson lines in abelian gauge theories arises from the vacuum polarization $\Pi(p^2)$ induced by massive charged matter. This causes a violation of the Ward identity, which, at weak coupling, can be computed at one loop from the Uehling correction to the Coulomb potential, as in~\cite{Cordova:2022rer}.

\section{Bad-Breaking Parameter in Diverse~Dimensions}\label{sec:bb_dd}
\begin{figure}[H]
  %  \centering
    \includegraphics[width=7.8cm, height=5.4cm]{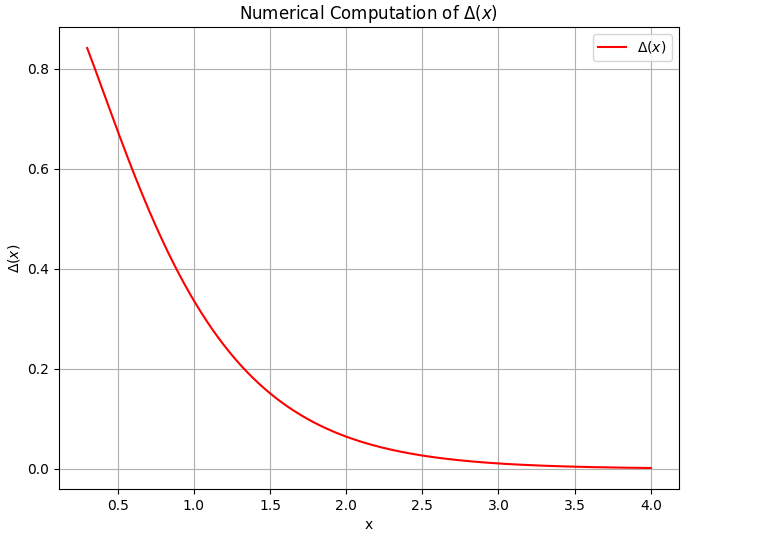}
    \caption{Bad-breaking parameter for a spinor in four~dimensions.}
    \label{fig:4dspinorgraph}
\end{figure}

In this section, we aim to derive the bad-breaking parameter in various dimensions and in a systematic fashion. We provide the detailed computation in four and five dimensions for both scalar and spinor QED, extending the estimates in~\cite{Cordova:2022rer}. From~textbook one-loop QED results, one can obtain the following result for the vacuum polarization $\Pi(p^{2})$ in the spinor case in $d$ spacetime dimensions:
\begin{eqaed}
    \Pi(p^2)= \frac{8e^2}{(4\pi)^{d/2}}\Gamma(2-\frac{d}{2})\int_{0}^{1}dx\, x(1-x)(m^2-p^{2}x(1-x))^{\frac{d}{2}-2}.
\end{eqaed}
In the scalar case, the~expression is slightly different, and~reads 
\begin{eqaed}
        \Pi(p^2)= \frac{2e^2}{(4\pi)^{d/2}}\Gamma(2-\frac{d}{2})\int_{0}^{1}dx\, x(2x-1)(m^2-p^{2}x(1-x))^{\frac{d}{2}-2}.
\end{eqaed}
In even dimensions, using dimensional regularization $d\to d-\epsilon$ leads to a renormalized expression in terms of the renormalized gauge coupling. In~four dimensions, this results in
\begin{eqaed}
    \Pi(p^2)= \frac{-e^2}{2\pi^2}\int_{0}^{1}dx\, x(1-x)\ln\left[1-\frac{p^2}{m^2}x(1-x)\right].
\end{eqaed}
This allows us to compute the one-loop correction to the Coulomb potential, which gives 
\begin{eqaed}
    \delta V(r) &= \frac{ie^2}{(2\pi)^2}\int_{-\infty}^{\infty}dp\, \frac{e^{ipr}}{pr}\Pi(p^2) \\
    & = \frac{ie^2}{(2\pi)^2}\frac{-e^2}{2\pi^2}\frac{(-2\pi i)}{6}\int_{2m}^{\infty}dp\, \frac{e^{-pr}}{pr}\sqrt{1-\frac{4m^2}{p^2}}\left(1+\frac{2m^2}{p^2}\right)\\
    & = \frac{-e^2}{4\pi r}\frac{e^2}{6\pi^2}\int_{2m}^{\infty}dp\, \frac{e^{-pr}}{p}\sqrt{1-\frac{4m^2}{p^2}}\left(1+\frac{2m^2}{p^2}\right).
\end{eqaed}
To obtain this result, we employed the relation
\begin{eqaed}
    F_{4}(pr) &\equiv \int d\Omega_{3}e^{ipr\cos\theta} \\
    & = \int_{0}^{2\pi}d\phi\int_{-1}^{+1}d\cos\theta\, e^{ipr\cos\theta}\\
    & = \frac{-2\pi i}{pr}(e^{ipr}- e^{-ipr}) \, ,
\end{eqaed}
{as well as the standard branch-cut properties of $\Pi(p^2)$. The~resulting integral ranges over momenta from $2m$ and above, reflecting the threshold for pair creation of charged particles participating in the screening.} Applying the modified Gauss law to this result, we obtain the desired effective charge:
\begin{eqaed}
    \delta q(r)=\frac{-e^{2}}{6\pi^{2}}\int_{2m}^{\infty}dp\,\frac{e^{-pr}}{p}(pr+1)\left(1+ \frac{2m^{2}}{p^{2}}\right)\sqrt{1-\frac{4m^{2}}{p^{2}}}.
\end{eqaed}
The above expressions encode the contribution of a single charged species of mass $m$ and charge $e$. The~complete vacuum polarization involves a sum over the spectrum, parametrized by masses $m_i$ and charges $n_i g$ as integer multiples of the gauge coupling $g$. Hence, the~(one-loop) bad-breaking parameter takes the form
\begin{eqaed}
        \delta_{\Lambda}= \sum_{i}\frac{n_{i}^{2}g^{2}}{6\pi^{2}}\int_{2m}^{\infty}dp\, pr^{2}e^{-pr}\left(1+ \frac{2m^2}{p^2}\right)\sqrt{1-\frac{4m^2}{p^2}}. 
\end{eqaed}
Recalling that the energy scale at which symmetry breaking is probed is just $\Lambda = r^{-1}$, we can simplify the above expression with the change of variables $s=p/m$, which yields
\begin{equation}
     \delta_{\Lambda}= \sum_{i}\frac{n_{i}^{2}g^{2}}{6\pi^{2}}\frac{m_{i}^2}{\Lambda^2}\int_{2}^{\infty}ds\, s\, e^{-sm_{i}/\Lambda}\left(1+ \frac{2}{s^2}\right)\sqrt{1-\frac{4}{s^2}},
\end{equation}
For the sake of brevity, we introduce
\begin{eqaed}
    \Delta_\text{4,f}(x)= x^2\int_{2}^{\infty}ds\, s\, e^{-sx}\left(1+ \frac{2}{s^2}\right)\sqrt{1-\frac{4}{s^2}} \, ,
\end{eqaed}
since it will appear often with $x = m/\Lambda$ in the ensuing discussions. The~final result can thus be recast as
\begin{eqaed}
    \delta_{\Lambda}= \sum_{i}\frac{n_{i}^{2}g^{2}}{6\pi^{2}}\Delta_\text{4,f}(x_{i}).
\end{eqaed}
The asymptotic behavior of $\Delta_\text{4,f}(x)$ for large and small values of $x$ is
\begin{eqaed}
    \Delta_\text{4,f}(x)\sim (1-2x^{2}+4x^{3})\quad \text{as}\quad x\to 0 \, ,\\
    \Delta_\text{4,f}(x)\sim \frac{\sqrt{x}}{e^{2x}}\quad \text{as}\quad x\to \infty \, .
\end{eqaed}
{\Cref{fig:4dspinorgraph} shows a numerical evaluation of $\Delta_{4,\text{f}}$, showing the expected exponential decay when symmetry breaking is probed at scales much smaller than the mass gap of charged~states.}\vspace{-6pt}

\unskip
\subsection{{{Scalar }
 QED in 4d}}
\begin{figure}[H]
 %   \centering
   \hspace{-0.5cm} \includegraphics[width=8cm, height=6.2cm]{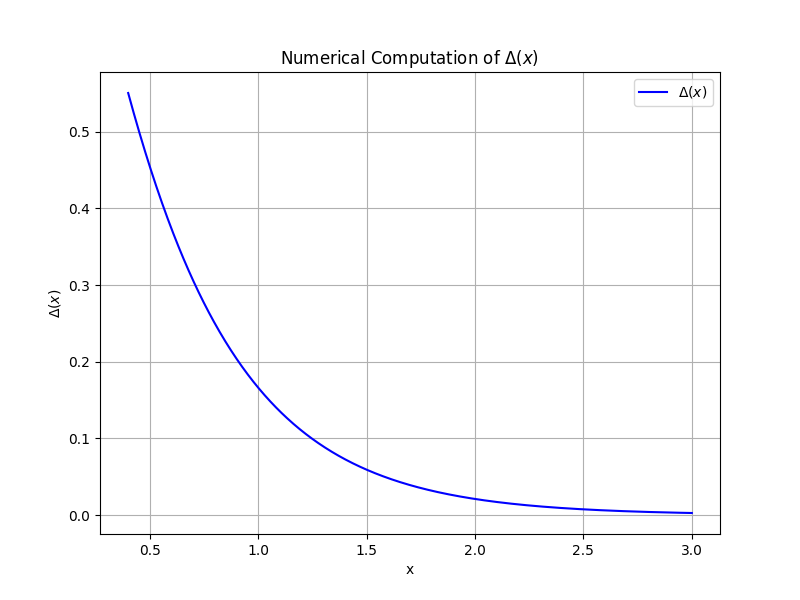}
    \caption{Bad-breaking parameter for a scalar field in four~dimensions.}
    \label{fig:4dscalargraph}
\end{figure}

For scalar QED in four dimensions, one should take the Fourier transform of
\begin{eqaed}
    V(p)=\frac{e^2}{p^2}\left(1-e^{2}\Pi(p^{2})\right),
\end{eqaed}
where $\Pi(p^2)$ now is 
\begin{eqaed}\label{eq:scalar_pi}
        \Pi(p^2)= \frac{-e^2}{8\pi^2}\int_{0}^{1}dx\, x(2x-1)\ln\left[1-\frac{p^2}{m^2}x(1-x)\right].
\end{eqaed}
The resulting one-loop correction to the potential is 
\begin{eqaed}
    \delta V(r)= \frac{ie^2}{(2\pi)^2}\int_{-\infty}^{\infty}dp\, \frac{e^{ipr}}{pr}\Pi(p^2)= \frac{-e^2}{4\pi r}\frac{e^2}{24\pi^2}\int_{2m}^{\infty}dp\, \frac{e^{-pr}}{p}\left(1-\frac{4m^2}{p^2}\right)^{\frac{3}{2}},
\end{eqaed}
where we use
\begin{equation}
    \int_{\frac{1}{2}- \frac{\beta}{2}}^{\frac{1}{2}+ \frac{\beta}{2}}dx\, x(2x-1)= \frac{1}{6}\left(1- \frac{4m^2}{p^2}\right)^{\frac{3}{2}},\quad \beta= \sqrt{1- \frac{4m^2}{p^2}}.
\end{equation}
The effective charge can be once more determined by taking a derivative with respect to $r$ and subsequently performing integration over $S^{2}$:
\begin{eqaed}
    \delta q(r)=\frac{-e^{2}}{24\pi^{2}}\int_{2m}^{\infty}dp\,\frac{e^{-pr}}{p}(pr+1)\left(1-\frac{4m^{2}}{p^{2}}\right)^{3/2}.
\end{eqaed}
{The factor of 24 results from the additional factor of 4 in the denominator of \cref{eq:scalar_pi} relative to the spinor case.} By applying the bad-breaking relation to the acquired effective charge, one arrives at
\begin{eqaed}
    \delta_{\Lambda}= \sum_{i}\frac{n_{i}^{2}g^{2}}{24\pi^{2}}\int_{2m}^{\infty}dp\, pr^{2}e^{-pr}\left(1-\frac{4m^2}{p^2}\right)^{\frac{3}{2}}.
\end{eqaed}
Performing the change of variables $s=p/m$ in the integral and writing $r=\frac{1}{\Lambda}$, as~in the case of spinor QED, the~above expression simplifies to
\begin{eqaed}
    \delta_{\Lambda}= \sum_{i}\frac{n_{i}^{2}g^{2}}{24\pi^{2}}\frac{m_{i}^2}{\Lambda^2}\int_{2}^{\infty}ds\, s\, e^{-sm_{i}/\Lambda}\left(1-\frac{4}{s^2}\right)^{\frac{3}{2}} \, .
\end{eqaed}
To write this expression in a more compact form, once again, we introduce an integral function, now for a four-dimensional scalar, according to
\begin{eqaed}
    \delta_{\Lambda} = \sum_{i}\frac{n_{i}^{2}g^{2}}{24\pi^{2}}\Delta_\text{4,b}(x_{i}) \, , \qquad \Delta_\text{4,b}(x)= x^2\int_{2}^{\infty}ds\, s\, e^{-sx}\left(1-\frac{4}{s^2}\right)^{\frac{3}{2}}.
\end{eqaed}
Similar to the spinor case, the~asymptotic behavior of $\Delta_\text{4,b}(x)$ is given by 
\begin{eqaed}
    \Delta_\text{4,b}(x) \sim (1-2x^{2}+4x^{3})\quad \text{as}\quad x\to 0 \, ,\\
    \Delta_\text{4,b}(x) \sim \frac{1}{\sqrt{x}e^{2x}}\quad \text{as}\quad x\to \infty \, .
\end{eqaed}
{\Cref{fig:4dscalargraph} shows a numerical evaluation of $\Delta_{4,\text{b}}$, showing once more the expected exponential decay displayed in \cref{fig:4dspinorgraph}.}
\unskip

\subsection{{{Spinor} QED in 5d}}
\begin{figure}[H]
  %  \centering
    \hspace{-0.3cm}\includegraphics[width=8cm, height=7cm]{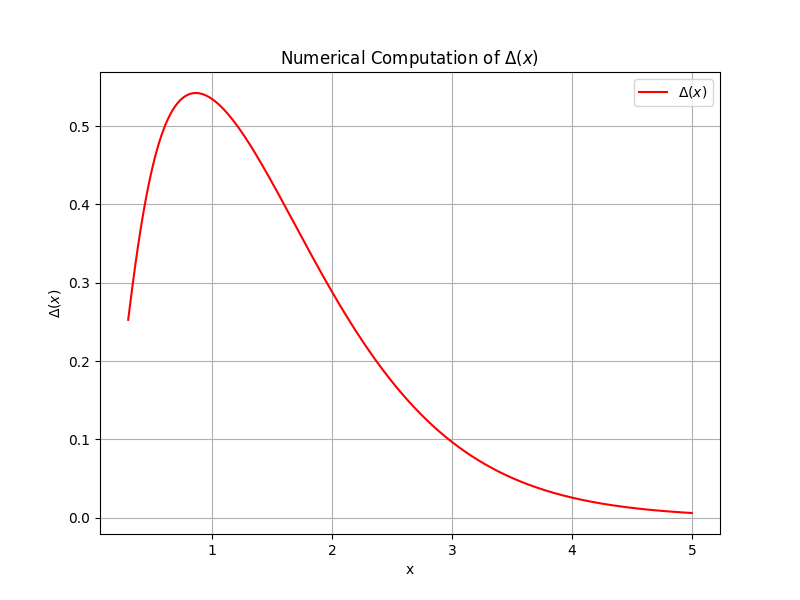}
    \caption{Bad-breaking parameter for a spinor in five dimensions. {In contrast with the four-dimensional case, the~plot shows a peak with a roughly twofold increase in $\Delta_{5,\text{f}}$ before the onset of large $x$ decay. Since bad breaking is a parametric notion, this does not have significant implications for our considerations.}}
    \label{fig:5dspinor}
\end{figure}

In five-dimensional spinor QED, $\Pi(p^{2})$ takes the form
\begin{eqaed}
    \Pi(p^2)=\frac{-e^2}{2\pi^2}\int_{0}^{1}dx\, x(1-x)(m^2-p^{2}x(1-x))^{1/2} \, ,
\end{eqaed}
which is finite as written. In~contrast to the 4d case, the~5d scenario presents technical complications; however, the~underlying logic and procedure remain unaltered. Writing the corrected potential in terms of a Fourier transform, one obtains
\begin{eqaed}
    \delta V(r)= \frac{-e^2}{(2\pi)^{2}r}\int_{0}^{\infty}dp\, \Pi(p^2)J_{1}(pr) \, ,
\end{eqaed}
where the Bessel function $J_1$ appears from the angular integrals according to
\begin{eqaed}
    \int_{0}^{2\pi}d\phi_{1}\int_{0}^{\pi}d\phi_{2}\sin\phi_{2}\int_{0}^{\pi}d\theta \sin^{2}\theta \,e^{ipr\cos \theta}= \frac{4\pi^{2}J_{1}(pr)}{pr} \, .
\end{eqaed}
To proceed further, we evaluate the integral over $x$, leading to
\begin{eqaed}
    \delta V(r)= \frac{e^{4} m}{(2\pi)^{2}(2\pi^2)}\int_{0}^{\infty}dp\, \frac{J_{1}(pr)}{r}\, W_{1}\left(\frac{p}{m}\right),
\end{eqaed}
where
\begin{eqaed}
    W_{1}(s)= \left(\frac{1}{8s^2}-\frac{1}{32}\right)+ \left(\frac{s}{32}+ \frac{1}{4s}+ \frac{1}{2s^3}\right)\left(\frac{\pi}{2}- \arcsin \left(\frac{s^2- 4}{s^2+ 4}\right)\right)+ \\ \left(\frac{s}{32}+ \frac{1}{8s}\right)\left(\frac{\pi}{2}+ \arcsin \left(\frac{s^2- 4}{s^2+ 4}\right)\right).
\end{eqaed}
As a result, the~five-dimensional effective charge is given by
\begin{eqaed}
    \delta q(r)= \frac{e^{2}m^{2}}{4\pi^2}\int_{0}^{\infty}\, W_{1}(s)\left((smr^{2})J_{1}^{'}(smr)- rJ_{1}(smr)\right).
\end{eqaed}
Applying the bad-breaking formula to a tower of charged particles thus yields
\begin{adjustwidth}{-\extralength}{0cm}
\begin{eqaed}
    \delta_{\Lambda}=\sum_{i} \frac{(g^{2}\Lambda)n_{i}^{2}}{4\pi^2} \frac{m_{i}^{2}}{\Lambda^{2}}\int_{0}^{\infty} ds\, W_{1}(s) \left( \left(\frac{sm_{i}}{\Lambda}\right)^{2} J_{1}^{''}\left(\frac{sm_{i}}{\Lambda}\right)  + \left(\frac{sm_{i}}{\Lambda}\right) J_{1}^{'}\left(\frac{sm_{i}}{\Lambda}\right)- J_{1}\left(\frac{sm_{i}}{\Lambda}\right) \right).
\end{eqaed}
\end{adjustwidth}
As before, introducing 
\begin{eqaed}
    \Delta_\text{5,f}(x)= x^{2}\int_{0}^{\infty}ds\, W_{1}(s) \left((sx)^{2} J_{1}^{''}(sx)+ (sx)J_{1}^{'}(sx)-J_{1}(sx)\right)
\end{eqaed}
simplifies the expression to
\begin{eqaed}
    \delta_{\Lambda}\approx \sum_{i} \frac{(g^{2}\Lambda)n_{i}^{2}}{4\pi^2}\Delta_\text{5,f}(x_{i}).
\end{eqaed}
{The function $\Delta_{5,\text{f}}$ is depicted in \cref{fig:5dspinor}. As in the four-dimensional case, this function decays for large arguments, as~expected from probing symmetry breaking by massive states at low energies. The~peak shown in \cref{fig:5dspinor} does not appear in four dimensions, but~it does not seem to carry physical significance: bad breaking is a parametric notion, while this peak amounts to roughly a factor of two variations before the onset of the large-argument~decay.}\vspace{-15pt}
\unskip
\subsection{{{Scalar} QED in 5d}}
\begin{figure}[H]
  %  \centering
   \hspace{-0.3cm} \includegraphics[width=8cm, height=7cm]{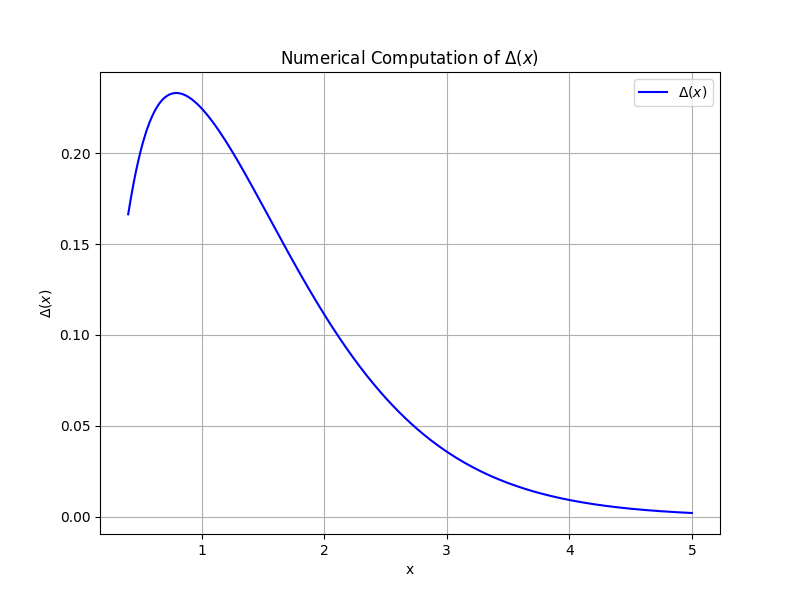}
    \caption{Bad-breaking parameter for a scalar field in five dimensions. {As in the spinor case, this plot shows a peak with a roughly twofold increase in $\Delta_{5,\text{b}}$ before the onset of large $x$ decay.}}
    \label{fig:5dscalar}
\end{figure}

Finally, we move on to five-dimensional scalar QED. After~performing the standard computation of one-loop vacuum polarization, we find
\begin{eqaed}
    \Pi(p^2)=\frac{-e^2}{8\pi^2}\int_{0}^{1}dx\, x(2x-1)(m^2-p^{2}x(1-x))^{1/2}.
\end{eqaed}
As before, we need to perform a Fourier transform of $\Pi(p^{2})$ to obtain the one-loop correction to the potential:
\begin{eqaed}
    \delta V(r)= \frac{-e^2}{(2\pi)^{2}r}\int_{0}^{\infty}dp\, \Pi(p^2)J_{1}(pr).
\end{eqaed}
Once again, in~order to derive an explicit formula for the one-loop potential, we perform the integral over $x$, leading to
\begin{eqaed}
    \delta V(r)= \frac{e^{4} m}{(2\pi)^{2}(8\pi^2)}\int_{0}^{\infty}dp\, \frac{J_{1}(pr)}{r}\, W_{2}\left(\frac{p}{m}\right),
\end{eqaed}
where $W_{2}(s)$
\begin{eqaed}
    W_{2}(s)= \left(\frac{1}{16}-\frac{1}{4s^2}\right)+ \left(\frac{s}{16}+ \frac{1}{2s}+ \frac{1}{s^3}\right)\left(\frac{\pi}{2}+ \arcsin \left(\frac{s^2-4}{s^2+ 4}\right)\right).
\end{eqaed}
Computing the effective charge from the modified Gauss law leads to
\begin{eqaed}
    \delta q(r)= \frac{e^{2}m^{2}}{16\pi^2}\int_{0}^{\infty}\, W_{2}(s)\left((smr^{2})J_{1}^{'}(smr)- rJ_{1}(smr)\right).
\end{eqaed}
Finally, the~bad-breaking parameter for a tower of charged particles takes again the usual form
\begin{eqaed}
    \delta_{\Lambda}\approx \sum_{i} \frac{(g^{2}\Lambda)n_{i}^{2}}{16\pi^2} \Delta_\text{5,b}(x_{i}), 
\end{eqaed}
with
\begin{eqaed}
    \Delta_\text{5,b}(x)= x^{2}\int_{0}^{\infty}ds\, W_{2}(s) \left((sx)^{2} J_{1}^{''}(sx)+ (sx)J_{1}^{'}(sx)-J_{1}(sx)\right).
\end{eqaed}
{This function is plotted in \cref{fig:5dscalar}. As~in the spinor case, there is a peak with no physical implications for the parametrics, as~well as the expected large-argument decay.}\vspace{-12pt}

\unskip
\subsection{{{Bad}-Breaking Parameter in Any Dimension}}
To conclude this section, we outline the key steps involved in deriving the bad-breaking parameter in an arbitrary spacetime dimension $d$. The~Fourier transform involving $\Pi(p^{2})$ to obtain the correction to the Coulomb potential takes the general form
\begin{eqaed}
    \delta V(r)&= -e^2\int \frac{d^{d-1}p}{(2\pi)^{d-1}}\frac{e^{ip.r}}{p^2}\Pi(p^2)\\
    &= \frac{-e^2}{(2\pi)^{d-1}}\int_{0}^{\infty}dp\, p^{d-4}\Pi(p^2)\int d\Omega_{d-1}e^{ipr\cos\theta}\\
    &= \frac{-e^2}{(2\pi)^{d-1}}\int_{0}^{\infty}dp\, p^{d-4}\Pi(p^2) F_{d}(pr),
\end{eqaed}
where we define the angular integral as
\begin{eqaed}
    F_{d}(pr)\equiv \int d\Omega_{d-1}e^{ipr\cos\theta}.
\end{eqaed}
To determine the effective charge in general dimension $d$, we use the modified Gauss in terms of a flux integral over $S^{d-2}$. This leads to
\begin{eqaed}
    q(r)= q_{\infty}\left[1- \frac{1}{(2\pi)^{d-1}}\int_{0}^{\infty}dp\, p^{d-4}\Pi(p^2)\int_{S^{d-2}}\frac{dF_{d}(pr)}{dr}\right].
\end{eqaed}
As a result, the~final formula for the bad-breaking parameter is
\begin{eqaed}
    \delta_{\Lambda}= \sum_i\frac{g^2 n_i^2}{(2\pi)^{d-1}}\int_{0}^{\infty}dp\, p^{d-4}\Pi(p^2)\int_{S^{d-2}}r\frac{d^{2}F_{d}(pr)}{dr^{2}}\, .
\end{eqaed}

\section{Bad-Breaking Parameter and BPS~States}\label{sec:bb_BPS}

In the preceding section, we introduced the technical machinery to compute bad-breaking parameters for various species of charged matter in diverse dimensions. As~discussed in \cref{sec:approx_symms}, the~intuition on badly broken symmetries in Quantum Gravity is motivated by bottom-up arguments. It is thus natural to seek simple, top-down constructions to assess the proposal. In~the context of string theory, Calabi--Yau compactifications of type II superstrings offer a rich and well-studied framework. Indeed, at~low energies, these settings are described by effective field theories with eight unbroken supercharges, namely, $\mathcal{N} = 2$ supergravity, whose multiplet content is determined by the topological properties of the internal manifold. Vector multiplets provide Abelian gauge sectors whose center one-form symmetry is broken by massive BPS states. The~latter arise from D-branes wrapping internal~cycles.

In this section, we aim to compute the bad-breaking parameter for these one-form symmetries in supergravity theories that arise, compactifying type IIA string theory and M-theory on a Calabi--Yau threefold $X$. This leads to eight supercharges in four and five dimensions, respectively. In~the appendices, we review some relevant material and notation. The~central charges of the BPS states that break the one-form symmetry are determined by the winding number of branes wrapped around each cycle and the volume of those cycles~\cite{Denef:2000nb, Denef:2002ru}. The~volume of each independent two-cycle (in string units) can be expressed as a function of K\"{a}hler moduli, $t^{i}$, with~$i = 1, \dots, h^{1,1}(X)$. Similarly, the~volumes of four-cycles are given by $\tau_i = C_{ijk} t^j t^k$, where $C_{ijk}$ indicates the triple intersection numbers of $X$. Finally, the~total volume of $X$ in string units is cubic in the $t^i$ and~yields the classical K\"{a}hler~potential
\begin{eqaed}
    \mathcal{V}=\frac{1}{6}C_{ijk}t^{i}t^{j}t^{k} \, , \qquad K=-\log 8\mathcal{V} \, .
\end{eqaed}
The central charges associated with BPS states are accompanied by the volumes of even-dimensional cycles according to
\begin{eqaed}
    Z_{6}= p^{0}\mathcal{V} \, ,\quad Z_{4}=p^{i}\tau_{i} \, ,\\
    Z_{2}= q_{i}t^{i} \, ,\quad Z_{0}=q_{0} \, .
\end{eqaed}
In the preceding section, we found that the parameter characterizing one-form symmetry breaking, which is induced by the screening effect, takes the general one-loop form:
\begin{eqaed}
    \delta_{\Lambda}\overset{g \ll 1}{\sim}\sum_{i}\frac{g^{2}n_{i}^{2}}{6\pi^{2}}\Delta\left(\frac{m_{i}}{\Lambda}\right).
\end{eqaed}
We begin from the four-dimensional case. The~complete one-form symmetry groups are $U(1)_\text{el}^{h^{1,1}(X)+1}$ and $U(1)_\text{m}^{h^{1,1}(X)+1}$ including the graviphoton. However, to~initially simplify matters, we shall begin evaluating the bad-breaking parameter for one factor at a time. For~the screening due to D2- and D4-branes wrapping two-cycles and four-cycles,\vspace{6pt}
\begin{eqaed}
    \delta_{\Lambda}^{U(1)_\text{el}^{(i)}}=\sum_{\{q_{j}\}}\frac{\left(g_\text{el}^{(i)}q_{i}\right)^{2}}{6\pi^{2}}\Delta\left(\frac{q_{j}t^{j}}{\Lambda}\right) \, , \\
    \delta_{\Lambda}^{U(1)_\text{m}^{(i)}}=\sum_{\{p^{j}\}}\frac{\left(g_\text{m}^{(i)}p^{i}\right)^{2}}{6\pi^{2}}\Delta\left(\frac{p^{j}\tau_{j}}{\Lambda}\right).
\end{eqaed}
Similarly, for~D6- and D0-branes,
\begin{eqaed}
    \delta_{\Lambda}& = \sum_{\{p^{0}\}}\frac{g_\text{m}^{2}(p^{0})^{2}}{6\pi^{2}}\Delta\left(\frac{\frac{p^{0}}{6}C_{ijk}t^{i}t^{j}t^{k}}{\Lambda}\right)=\sum_{\{p^{0}\}}\frac{g_\text{m}^{2}(p^{0})^{2}}{6\pi^{2}}\Delta\left(\frac{\frac{p^{0}}{8}e^{-K}}{\Lambda}\right) , \\
    \delta_{\Lambda}&= \sum_{\{q_{0}\}}\frac{g^{2}q_{0}^{2}}{6\pi^{2}}\Delta\left(\frac{q_{0}}{\Lambda}\right).
\end{eqaed}
In the above expressions, as~well as in the following, $\Lambda$ is expressed in string units, as~per the conventions outlined above. One may also combine the screening effects due to multi-charge states to obtain a more precise result. The~resulting expressions for electric and magnetic bad-breaking parameters,
\begin{eqaed}
    \delta_{\Lambda}\sim \sum_{\{q_{0}, q_{j}\}}\frac{g_\text{el}^{2}(q_{0}^{2}+ q_{j}^{2})}{6\pi^{2}}\Delta\left(\frac{q_{0}+q_{j}t^{j}}{\Lambda}\right),\\
    \delta_{\Lambda}\sim \sum_{\{p^{0}, p^{j}\}}\frac{g_\text{m}^{2}\left((p^{0})^{2}+ (p^{j})^{2}\right)}{6\pi^{2}}\Delta\left(\frac{\frac{p^{0}}{8}e^{-K}+ \frac{p^{j}}{2}C_{jlm}t^{l}t^{\text{m}}}{\Lambda}\right),
\end{eqaed}
are still incomplete, as~we shall address later on when accounting for mixing between different $U(1)$ factors. {Strictly speaking, these sums are finite, ranging over states with masses not exceeding the species scale. However, in the~infinite-distance limits in which we are interested, an infinitely number of states participate in the sum. In~the examples we shall discuss in the following, these infinite sums arrange into convergent, geometric series due to the finite threshold to pair production in the loop integral.}

We will now proceed to describe the above sums in more detail. Consider a single $U(1)$ factor at a time when the index $i$ is fixed and~the ensuing simplifications depend on whether the terms in the sums correspond with $i$. Thus, we divide the sums over BPS charges into three cases for the indices. As~a result, we arrive at the rather extensive expression
\begin{eqaed}
    \sum_{\{q_{j}\}}\sum_{a,b\neq i}\frac{g_\text{el}^{2}q_{i}^{2}}{6\pi^{2}}\frac{q_{a}q_{b}t^{a}t^{b}}{\Lambda^{2}}\int_{2}^{\infty}ds\, s e^{\frac{-sq_{j}t^{j}}{\Lambda}}\sqrt{1-\frac{4}{s^{2}}}\left(1+\frac{2}{s^{2}}\right)\\
    +2\sum_{\{q_{j}\}}\sum_{a\neq i}\frac{g_\text{el}^{2}q_{i}^{3}}{6\pi^{2}}\frac{t^{i}q_{a}t^{a}}{\Lambda^{2}}\int_{2}^{\infty}ds\, s e^{\frac{-sq_{j}t^{j}}{\Lambda}}\sqrt{1-\frac{4}{s^{2}}}\left(1+\frac{2}{s^{2}}\right)\\
    + \sum_{\{q_{j}\}}\frac{g_\text{el}^{2}q_{i}^{4}(t^{i})^{2}}{6\pi^{2}\Lambda^{2}}\int_{2}^{\infty}ds\, s e^{\frac{-sq_{j}t^{j}}{\Lambda}}\sqrt{1-\frac{4}{s^{2}}}\left(1+\frac{2}{s^{2}}\right).
\end{eqaed}
Here, we used the explicit form $\Delta(x) = \Delta_\text{4,f}(x)$ for definiteness. However, our strategy to recast the sums in a way that is amenable to asymptotic analyses is independent of this choice, insofar as $\Delta(x)$ has the same qualitative form in terms of a Schwinger-like integral. In~a parallel fashion, one can derive an expression for magnetic one-form symmetry breaking that exhibits a similar structure:\vspace{6pt}
\begin{eqaed}
    \sum_{\{p^{j}\}}\sum_{a,b\neq i}\frac{g_\text{m}^{2}(p^{i})^{2}}{6\pi^{2}}\frac{p^{a}p^{b}\tau_{a}\tau_{b}}{\Lambda^{2}}\int_{2}^{\infty}ds\, s e^{\frac{-sp^{j}\tau_{j}}{\Lambda}}\sqrt{1-\frac{4}{s^{2}}}\left(1+\frac{2}{s^{2}}\right)\\
    + 2\sum_{\{p^{j}\}}\sum_{a\neq i}\frac{g_\text{m}^{2}(p^{i})^{3}}{6\pi^{2}}\frac{\tau_{i} p^{a}\tau_{a}}{\Lambda^{2}}\int_{2}^{\infty}ds\, s e^{\frac{-sp^{j}\tau_{j}}{\Lambda}}\sqrt{1-\frac{4}{s^{2}}}\left(1+\frac{2}{s^{2}}\right)\\
    + \sum_{\{p^{j}\}}\frac{g_\text{m}^{2}(p^{i})^{4}}{6\pi^{2}}\frac{(\tau_{i})^{2}}{\Lambda^{2}}\int_{2}^{\infty}ds\, s e^{\frac{-sp^{j}\tau_{j}}{\Lambda}}\sqrt{1-\frac{4}{s^{2}}}\left(1+\frac{2}{s^{2}}\right).
\end{eqaed}

{Simplifying assumptions:}
In order to simplify the above expressions, we can make some simplifying assumptions and borrow some standard techniques from grand-canonical computations in statistical mechanics. To~begin with, we reassemble the sums over tuples of charges into products of single sums, obtaining
\begin{adjustwidth}{-\extralength}{0cm}
\begin{eqaed}
    \frac{g_\text{el}^{2}(t^{i})^{2}}{6\pi^{2}\Lambda^{2}}\int_{2}^{\infty}ds\, s \sqrt{1-\frac{4}{s^{2}}}\left(1+\frac{2}{s^{2}}\right)\left(\sum_{\{q_{i}\}}q_{i}^{4}e^{\frac{-sq_{i}t^{i}}{\Lambda}}\right)\left(\prod_{j\neq i}\sum_{\{q_{j}\}}e^{\frac{-sq_{j}t^{j}}{\Lambda}}\right)+\\
    \sum_{a\neq i}\frac{2g_\text{el}^{2}t^{i}t^{a}}{6\pi^{2}\Lambda^{2}}\int_{2}^{\infty}ds\, s \sqrt{1-\frac{4}{s^{2}}}\left(1+\frac{2}{s^{2}}\right)\left(\sum_{\{q_{i}\}}q_{i}^{3}e^{\frac{-sq_{i}t^{i}}{\Lambda}}\right)\left(\sum_{\{q_{a}\}}q_{a}e^{\frac{-sq_{a}t^{a}}{\Lambda}}\right)\left(\prod_{j\neq i, a}\sum_{\{q_{j}\}}e^{\frac{-sq_{j}t^{j}}{\Lambda}}\right)\\
    + \sum_{a\neq i}\frac{g_\text{el}^{2}(t^{a})^{2}}{6\pi^{2}\Lambda^{2}}\int_{2}^{\infty}ds\, s \sqrt{1-\frac{4}{s^{2}}}\left(1+\frac{2}{s^{2}}\right)\left(\sum_{\{q_{i}\}}q_{i}^{2}e^{\frac{-sq_{i}t^{i}}{\Lambda}}\right)\left(\sum_{\{q_{a}\}}q_{a}^{2}e^{\frac{-sq_{a}t^{a}}{\Lambda}}\right)\left(\prod_{j\neq i, a}\sum_{\{q_{j}\}}e^{\frac{-sq_{j}t^{j}}{\Lambda}}\right)\\
    + \sum_{a\neq b\neq i}\frac{g_\text{el}^{2}t^{a}t^{b}}{6\pi^{2}\Lambda^{2}}\int_{2}^{\infty}ds\, s \sqrt{1-\frac{4}{s^{2}}}\left(1+\frac{2}{s^{2}}\right)\left(\sum_{\{q_{i}\}}q_{i}^{2}e^{\frac{-sq_{i}t^{i}}{\Lambda}}\right) \left(\sum_{\{q_{a}\}}q_{a}e^{\frac{-sq_{a}t^{a}}{\Lambda}}\right) \left(\sum_{\{q_{b}\}}q_{b}e^{\frac{-sq_{b}t^{b}}{\Lambda}}\right)\\ \left(\prod_{j\neq i, a, b}\sum_{\{q_{j}\}}e^{\frac{-sq_{j}t^{j}}{\Lambda}}\right)
\end{eqaed}
\end{adjustwidth}
analogously for the magnetic case. In~order to present the expression in a more concise form, we introduce the function $f_{m}^{k}(x)=\sum_{\{q_{k}\}}q_{k}^{m}e^{-q_{k}x}$. This simplifies the above to
\begin{eqaed}
    \frac{g_\text{el}^{2}(t^{i})^{2}}{6\pi^{2}\Lambda^{2}}\int_{2}^{\infty}ds\, s \sqrt{1-\frac{4}{s^{2}}}\left(1+\frac{2}{s^{2}}\right) f_{4}^{i}\left(\frac{st^{i}}{\Lambda}\right)\prod_{j\neq i}f_{0}^{j}\left(\frac{st^{j}}{\Lambda}\right)+\\
    \sum_{a\neq i}\frac{2g_\text{el}^{2}t^{i}t^{a}}{6\pi^{2}\Lambda^{2}}\int_{2}^{\infty}ds\, s \sqrt{1-\frac{4}{s^{2}}}\left(1+\frac{2}{s^{2}}\right)f_{3}^{i}\left(\frac{st^{i}}{\Lambda}\right)f_{1}^{a}\left(\frac{st^{a}}{\Lambda}\right)\prod_{j\neq i, a}f_{0}^{j}\left(\frac{st^{j}}{\Lambda}\right)+\\
    \sum_{a\neq i}\frac{g_\text{el}^{2}(t^{a})^{2}}{6\pi^{2}\Lambda^{2}}\int_{2}^{\infty}ds\, s \sqrt{1-\frac{4}{s^{2}}}\left(1+\frac{2}{s^{2}}\right)f_{2}^{i}\left(\frac{st^{i}}{\Lambda}\right)f_{2}^{a}\left(\frac{st^{a}}{\Lambda}\right)\prod_{j\neq i, a}f_{0}^{j}\left(\frac{st^{j}}{\Lambda}\right)+\\
    \sum_{a\neq b\neq i}\frac{g_\text{el}^{2}t^{a}t^{b}}{6\pi^{2}\Lambda^{2}}\int_{2}^{\infty}ds\, s \sqrt{1-\frac{4}{s^{2}}}\left(1+\frac{2}{s^{2}}\right)f_{2}^{i}\left(\frac{st^{i}}{\Lambda}\right)f_{1}^{a}\left(\frac{st^{a}}{\Lambda}\right)f_{1}^{b}\left(\frac{st^{b}}{\Lambda}\right)\prod_{j\neq i, a, b}f_{0}^{j}\left(\frac{st^{j}}{\Lambda}\right).
\end{eqaed}
Each sum now has a fixed number of instances of $q_i$ and~can be further simplified by replacing multiplication by $q^a$ with an appropriate derivative $\partial_{t^a}$, which we can then factor out of the sum. For~example, we can express the last term via the combination $\partial_{t^{i}} \partial_{t^{i}} \partial_{t^{a}} \partial_{t^{b}}$ preceding the integral, allowing us to recast it according to\vspace{6pt}
\begin{eqaed}
    \sum_{a\neq b\neq i}\frac{g_\text{el}^{2}t^{a}t^{b}\Lambda^{2}} {6\pi^{2}} \partial_{t^{i}} \partial_{t^{i}} \partial_{t^{a}} \partial_{t^{b}}\int_{2}^{\infty}ds\, \frac{1}{s^{3}} \sqrt{1-\frac{4}{s^{2}}}\left(1+\frac{2}{s^{2}}\right) \left(\sum_{\{q_{i}\}}e^{\frac{-sq_{i}t^{i}}{\Lambda}}\right) \left(\sum_{\{q_{a}\}}e^{\frac{-sq_{a}t^{a}}{\Lambda}}\right)\\ \left(\sum_{\{q_{b}\}}e^{\frac{-sq_{b}t^{b}}{\Lambda}}\right) \left(\prod_{j\neq i, a, b}\sum_{\{q_{j}\}}e^{\frac{-sq_{j}t^{j}}{\Lambda}}\right) .
\end{eqaed}

Thus far, the~only simplifying assumption has been the restriction to a single $U(1)$ factor at a time. To~make further progress, from~now, we simplify the treatment of the degeneracies of BPS charges, which are implicit in the sum. When estimating degeneracies of BPS states that become light in infinite-distance limits (thus dominating the above sums) with index-like quantities, such as Gopakumar--Vafa invariants, it has been observed in large classes of examples that two types of behaviors arise. Namely, the~(estimated) degeneracies either remain roughly constant or they grow exponentially~\cite{Rudelius:2023odg}. The~former commonly arises in circle compactifications, while the latter arises when string excitations become light in Planck units. This pattern is consistent with the emergent string conjecture~\cite{Lee:2018urn, Lee:2019wij, Lee:2019xtm}. Therefore, in~the following, we consider the simplest versions of these behaviors: a strict exponential growth in the charges would behave as a shift in the moduli, at~least qualitatively, since the Schwinger integrals begin as $s=2$. Hence, we simply consider constant degeneracies, which allow us to evaluate the infinite sums in closed form. Finally, BPS charges span cones determined by the effective divisors of the internal manifold. As~such, they may not afford a simplicial parametrization in terms of tuples of positive integers in some basis. Since we evaluate the asymptotic behavior of bad-breaking parameters for the case of a single modulus, where there is only one electric or magnetic charge to sum over, we also restrict ourselves to unconstrained charge lattices in the positive orthant $\mathbb{Z}^{h^{1,1}(X)}_{+}$, namely, tuples of positive integers ({{we} thank N. Gendler for insightful correspondence on these matters}). We leave a more detailed and systematic numerical study of the infinite sums with full BPS degeneracies to future~work.

As mentioned above, the~infinite sums over charges become geometric series. As~a result, we obtain the much simpler expression
\begin{eqaed}
    \delta_{\Lambda}^{\text{el}}= \sum_{a,b}\frac{g_\text{el}^{2}t^{a}t^{b}\Lambda^{2}}{6\pi^{2}}\partial_{t^{i}} \partial_{t^{i}} \partial_{t^{a}} \partial_{t^{b}}\int_{2}^{\infty}ds\, \frac{1}{s^{3}} \sqrt{1-\frac{4}{s^{2}}}\left(1+\frac{2}{s^{2}}\right)\left(\prod_{n=1}^{h^{1,1}(X)}\frac{1}{1-e^{\frac{-st^{n}}{\Lambda}}}\right).
\end{eqaed}
Similarly, for~the magnetic one-form symmetry, the~complicated expression
\begin{eqaed}
    \sum_{a\neq i}\frac{2g_\text{m}^{2}\tau_{i}\tau_{a}\Lambda^{2}}{6\pi^{2}}\partial_{\tau_{i}}\partial_{\tau_{i}}\partial_{\tau_{i}}\partial_{\tau_{a}}\int_{2}^{\infty}\frac{ds}{s^{3}}\left(1+ \frac{2}{s^{2}}\right)\sqrt{1-\frac{4}{s^{2}}}g_{0}^{i}\left(\frac{s\tau_{i}}{\Lambda}\right)g_{0}^{a}\left(\frac{s\tau_{a}}{\Lambda}\right)\prod_{j\neq i, a}g_{0}^{j}\left(\frac{s\tau_{j}}{\Lambda}\right)+\\
    \sum_{a\neq i}\frac{g_\text{m}^{2}\tau_{a}^{2}\Lambda^{2}}{6\pi^{2}}\partial_{\tau_{i}}\partial_{\tau_{i}}\partial_{\tau_{a}}\partial_{\tau_{a}}\int_{2}^{\infty}\frac{ds}{s^{3}}\left(1+ \frac{2}{s^{2}}\right)\sqrt{1-\frac{4}{s^{2}}}g_{0}^{i}\left(\frac{s\tau_{i}}{\Lambda}\right)g_{0}^{a}\left(\frac{s\tau_{a}}{\Lambda}\right)\prod_{j\neq i, a}g_{0}^{j}\left(\frac{s\tau_{j}}{\Lambda}\right)\\
    +\text{remaining terms},
\end{eqaed}
where $g_{m}^{k}(x)= \sum_{\{p^{i}\}}(p^{k})^{m}e^{-p^{k}x}$ serves as a counterpart to $f_{m}^{k}(x)$, simplifies dramatically~to
\begin{eqaed}
    \delta_{\Lambda}^{\text{m}}=\sum_{a,b}\frac{g_\text{m}^{2}\tau_{a}\tau_{b}\Lambda^{2}}{6\pi^{2}}\partial_{\tau_{i}} \partial_{\tau_{i}} \partial_{\tau_{a}} \partial_{\tau_{b}}\int_{2}^{\infty}ds\, \frac{1}{s^{3}} \sqrt{1-\frac{4}{s^{2}}}\left(1+\frac{2}{s^{2}}\right)\left(\prod_{n=1}^{h^{1,1}(X)}\frac{1}{1-e^{\frac{-s\tau_{n}}{\Lambda}}}\right).
\end{eqaed}

\subsection{{{Including}
Mixing between Abelian Factors}}

Despite the simplifying assumptions we make to obtain these results, it is feasible to refine our treatment of bad breaking of one-form symmetries to include the full $U(1)^{h^{1,1}(X)}_\text{el}$ and $U(1)^{h^{1,1}(X)}_\text{m}$ symmetries arising from vector multiplets. This requires slightly modifying the definition of the bad-breaking parameter since in the chosen basis of BPS charges, the gauge couplings are generically mixed via kinetic terms of the form $f_{IJ}(t)F^{I}\wedge \star F^{J}$. As~a result, the~effective one-form charge carried by Wilson lines arising from one-loop vacuum polarization takes the schematic form
\begin{eqaed}
    \delta Q_{i}(r)= \sum_{\{q\}} Q_{k}^{\infty}f^{jk}q_{i}q_{j}\left(\text{1-loop term}\right),
\end{eqaed}
where $Q_{\infty}$ denotes the IR value of the effective charge, and~analogously for 't Hooft lines. In~light of these considerations, it is clear that the extent of symmetry breaking is more generally quantified by a \emph{bad-breaking matrix}, whose components in our case read
\begin{eqaed}\label{sec:el_bb_simple}
    (\delta_{\Lambda}^{\text{el}})_{i}^{j}= \frac{\Lambda^{2}}{6\pi^{2}}f^{jk}t^{a}t^{b}\partial_{t^{i}} \partial_{t^{k}} \partial_{t^{a}} \partial_{t^{b}}\int_{2}^{\infty}ds\, \frac{1}{s^{3}} \sqrt{1-\frac{4}{s^{2}}}\left(1+\frac{2}{s^{2}}\right)\left(\prod_{n=1}^{h^{1,1}(Y)}\frac{1}{1-e^{\frac{-st^{n}}{\Lambda}}}\right).
\end{eqaed}
This is a straightforward generalization of the formulae presented above. Similarly, the~one-loop bad-breaking matrix for magnetic one-form symmetry breaking is given by
\begin{eqaed}\label{eq:mag_bb_simple}
    (\delta_{\Lambda}^{\text{m}})_{j}^{i}= \frac{\Lambda^{2}}{6\pi^{2}}f_{jk}\tau_{a}\tau_{b}\partial_{\tau_{i}} \partial_{\tau_{k}} \partial_{\tau_{a}} \partial_{\tau_{b}}\int_{2}^{\infty}ds\, \frac{1}{s^{3}} \sqrt{1-\frac{4}{s^{2}}}\left(1+\frac{2}{s^{2}}\right)\left(\prod_{n=1}^{h^{1,1}(Y)}\frac{1}{1-e^{\frac{-s\tau_{n}}{\Lambda}}}\right).
\end{eqaed}
In this context, the~eigenvalues of bad-breaking matrices encode the bad-breaking parameters associated with unmixed~symmetries.

\subsection{{{Consequences}for a Single Modulus}}

What are the consequences of imposing the bad-breaking requirement $\Lambda_\text{bb} \lesssim M_\text{Pl}$ or $\Lambda_\text{bb} \lesssim \Lambda_\text{sp}$ in the supergravity settings we have examined? Addressing this important question is the primary motivation of this section. In~order to address this, we need to compute the asymptotic behavior of the bad-breaking parameters, species scale, gauge couplings, and~their mixing matrices (in more intricate scenarios) as functions of moduli. This is the central question we aim to~explore.

In order to derive concrete expressions to this end, here, we focus on models with a single (K\"{a}hler) modulus. {{Notice} that the asymptotics of models with a single modulus do not necessarily reflect those of models with multiple moduli where one is sent to an infinite distance. This is due to the fact that additional terms accounting for extra moduli can dominate, as~pointed out, e.g.,~in~\cite{vandeHeisteeg:2023dlw} when studying the slope of the species scale}, which greatly simplifies matters. We can construct explicit examples using tools such as \texttt{{CYTools}
}~\cite{Demirtas:2022hqf}. The~upshot is that the more general expressions in Equations \eqref{sec:el_bb_simple} and \eqref{eq:mag_bb_simple} further simplify to
\begin{eqaed}
    \delta_{\Lambda}^{\text{el}} & = \frac{g^{2}t^{2}\Lambda^{2}}{6\pi^{2}}\partial_{t}^{4}\int_{2}^{\infty}ds\, \frac{1}{s^{3}} \sqrt{1-\frac{4}{s^{2}}}\left(1+\frac{2}{s^{2}}\right)\left(\frac{1}{1-e^{\frac{-st}{\Lambda}}}\right) , \\
    \delta_{\Lambda}^{\text{m}}& = \frac{g^{2}C^{2}t^{4}\Lambda^{2}}{6\pi^{2}}\partial_{\tau}^{4}\int_{2}^{\infty}ds\, \frac{1}{s^{3}} \sqrt{1-\frac{4}{s^{2}}}\left(1+\frac{2}{s^{2}}\right)\left(\frac{1}{1-e^{\frac{-s\tau}{\Lambda}}}\right).    
\end{eqaed}

In order to investigate the bad-breaking condition, we now study the small-modulus limit $t\ll \Lambda$ and the large-modulus limit $t \gg \Lambda$. Let us remark that despite this nomenclature, when studying the bad-breaking parameter at the species scale $\Lambda = \Lambda_\text{sp}(t)$ {a priori}, the limit $t \ll \Lambda$ could be achieved for $t \ll 1$, insofar as the $t$-dependence of $\Lambda_\text{sp}$ allows. For~the time being, we keep $\Lambda$ constant, and~later include its $t$-dependence. 

\textbf{{Small Modulus:}
} Expanding the integrand, we find
\begin{eqaed}
    & g^{2}\Lambda^{2}t^{2}\partial_{t}^{4}\int_{2}^{\infty}ds\, \frac{1}{s^{3}} \sqrt{1-\frac{4}{s^{2}}}\left(1+\frac{2}{s^{2}}\right)\left(1-(1-st/\Lambda)\right)^{-1}\\
    & \sim g^{2}\Lambda^{2}t^{2}\partial_{t}^{4}\int_{2}^{\infty}ds\, \frac{1}{s^{3}} \sqrt{1-\frac{4}{s^{2}}}\left(1+\frac{2}{s^{2}}\right)\left(\frac{\Lambda}{st}\right)\\
    & \sim \frac{g^{2}\Lambda^{3}}{t^{3}}\int_{2}^{\infty}\frac{ds}{s^{4}} \sqrt{1-\frac{4}{s^{2}}}\left(1+\frac{2}{s^{2}}\right)\\
    & \Longrightarrow \delta_{\Lambda}^{\text{el}}\sim \frac{g^{2}\Lambda^{3}}{t^{3}}.
\end{eqaed}

\textbf{{Large Modulus:}} In this case, one can perform a saddle-point approximation of the integral. Since the slowest exponential decay rate appearing in the integrand is achieved by the infimum of the integration range $s=2$, performing the changes of variable $r=s-2$ and $r=u^2$, we find
\begin{eqaed}
    & g^{2}\Lambda^{2}t^{2}\partial_{t}^{4}\int_{2}^{\infty}ds\, \frac{1}{s^{3}} \sqrt{1-\frac{4}{s^{2}}}\left(1+\frac{2}{s^{2}}\right)\left(1+ e^{\frac{-st}{\Lambda}}\right)\\
    & \sim g^{2}\Lambda^{2}t^{2}\partial_{t}^{4}\int_{2}^{\infty}ds\, \frac{1}{s^{3}} \sqrt{1-\frac{4}{s^{2}}}\left(1+\frac{2}{s^{2}}\right)e^{\frac{-st}{\Lambda}}\\
    & \sim \frac{g^{2}t^{2}}{\Lambda^{2}}e^{\frac{-2t}{\Lambda}}\int_{0}^{\infty}dr\, \sqrt{r}e^{\frac{-tr}{\Lambda}}\\
    & \sim \frac{g^{2}t^{2}}{\Lambda}e^{\frac{-2t}{\Lambda}}\partial_{t}\int_{0}^{\infty}du\, e^{\frac{-tu^{2}}{\Lambda}}\\
    & \Longrightarrow \delta_{\Lambda}^{\text{el}}\sim \frac{g^{2}t^{1/2}}{\Lambda^{1/2}}e^{\frac{-2t}{\Lambda}}.
\end{eqaed}
In the above expression, $\Lambda$ is kept constant in string units. However, from~the point of view of effective field theory, it is more natural to keep $\Lambda$ constant in Planck units, which amounts to replacing $t$ with $t \, e^{K/2}$. Since the K\"{a}hler potential is the logarithm of a cubic function, up~to a rescaling of $\Lambda$ by numerical prefactors, one can effectively substitute $t$ with $t^{-1/2}$. This results in an interchange of the two limiting~behaviors. 

\textbf{{Small Modulus:}}
\begin{eqaed}
    \delta_{\Lambda}^{\text{el}}\sim \frac{g^{2} t^{11/4}}{\Lambda^{1/2}} e^{-\frac{2}{\Lambda \sqrt{t}}}.
\end{eqaed}

\textbf{{Large Modulus:}}
\begin{eqaed}
    \delta_{\Lambda}^{\text{el}}\sim \frac{g^{2}\Lambda^3}{t^{3/2}}.
\end{eqaed}

Analogous results hold for the magnetic case. The~physical interpretation of the results presented in~\cite{Rudelius:2023odg} indicates that when the states screening the Wilson lines are heavy, the~breaking parameter remains small. In~other words, the~effect of the breaking is minimal when assessed at scales considerably lower than the charged mass gap. As~the states become light, the~breaking parameter becomes significant, leading to a severe breaking of~symmetry. 

Notably, we observe that in the large-$t$ infinite-distance limit, the~lightest tower of BPS states are D0-branes, whose masses scale like $\frac{q_0}{t^{3/2}}$ in Planck units. However, these states are not charged under the vector multiplets; wrapped D2-branes instead have masses scaling like $\frac{q_1 t}{t^{3/2}}$. The~associated species scale is given by $\Lambda_{sp} \sim t^{-1/6}$. For~large $t$, the combination $\Lambda_{sp}(t) \sqrt{t}$ is large, allowing the large-modulus asymptotics of the bad-breaking parameter to be consistent. Since, in this limit, the gauge coupling scales as $g^2 \sim 1/t$, we find $\delta_{\Lambda_{sp}} \sim t^{-3}$. Therefore, in~this toy model example, the electric one-form symmetry is not badly broken at the species scale defined by charged BPS states. On~the one hand, this may not be problematic, since the species scale associated with BPS states does not necessarily coincide with the full species scale~\cite{Long:2021jlv}. On~the other hand, the~bad-breaking parameter seems also to remain at the Planck scale. It would be fascinating to incorporate more moduli and/or study detailed, top-down scenarios to further explore the connection (if any) between bad breaking and the species scale, and~the bad-breaking condition more broadly. Moreover, it is important to note that the magnetic weak-gravity cutoff $\Lambda_{U(1)} = g M_\text{Pl} \sim t^{-1/2}$ \cite{Arkani-Hamed:2006emk, Harlow:2022ich} acts as a transition scale in these expressions, as~the dependence on $t$ in the integrals and prefactors arises in the form $\Lambda \sqrt{t}$. However, the~leftover dependence in the gauge coupling still makes the overall expression small in the large-$t$ limit, consistent with the above considerations. At~any rate, the~important role of $\Lambda_{U(1)}$ in this context resonates with the considerations of~\cite{Heidenreich:2017sim}.

As a final remark before addressing the five-dimensional case, the~computation for a single modulus was performed by considering contributions from only one type of wrapping of internal cycles, specifically D2-branes. As~mentioned earlier, it is possible to combine the effects of BPS charges from different types of cycles. For~instance, including D0--D2 and D4--D6 bound states, the bad-breaking parameters for of a single modulus should be refined according to
\begin{eqaed}
    \delta_{\Lambda}^{\text{el}}&= \sum_{\{q_{0}, q_{1}\}}\frac{g_\text{el}^{2}\left((q_{0})^{2}+ (q_{1})^{2}\right)}{6\pi^{2}}\Delta\left(\frac{q_{0}+ q_{1}t}{\Lambda}\right)\\
    \delta_{\Lambda}^{\text{m}}&= \sum_{\{p^{0}, p^{1}\}}\frac{g_\text{m}^{2}\left((p^{0})^{2}+ (p^{1})^{2}\right)}{6\pi^{2}}\Delta\left(\frac{\frac{p^{0}}{8}e^{-K}+ \frac{p^{1}}{2}Ct^{2}}{\Lambda}\right).
\end{eqaed}
For example, in~the magnetic case, the~aforementioned results follow in the (putative) limit of a large intersection number $C \gg 1$, where the second term in the numerator of $\Delta$ dominates. Conversely, in~the large-volume limit, the~first term in the numerator takes precedence, in~which case, the bad-breaking parameter would be instead approximated by
\begin{eqaed}
    \delta_{\Lambda}^{\text{m}}\sim \sum_{\{p^{0}\}}\frac{g_\text{m}^{2}\left(p^{0}\right)^{2}}{6\pi^{2}}\Delta\left(\frac{\frac{p^{0}}{8}e^{-K}}{\Lambda}\right).
\end{eqaed}

\subsection{{{Bad}-Breaking Parameter in 5d Supergravity}}

Let us now proceed to the five-dimensional case. The~geometry of vector multiplets in 5$d$ supergravity and~their low-energy couplings are described by a cubic prepotential~\cite{DallAgata:2021uvl}:
\begin{eqaed}
    \mathcal{F}=\frac{1}{6}C_{IJK}Y^{I}Y^{J}Y^{K} \, .
\end{eqaed}
In the settings we consider in this paper, these effective theories arise, compactifying M-theory on a Calabi--Yau threefold $X$ \cite{Cadavid:1995bk, Ferrara:1996hh}. {{It} is worth noting that asymmetric orbifold techniques can produce perturbative string models that are not of this type since they lack hypermultiplets~\cite{Baykara:2023plc}}, in~which case, the moduli $Y^{I}$, $I=0, \dots , h^{1,1}(X)-1$ represent the volume of a basis of two-cycles and $C_{IJK}$ are triple intersection~numbers.

Following the conventions of~\cite{Rudelius:2023odg}, the~mass of electric BPS particles in Planck units is given by
\begin{eqaed}
    m(q_{I})=\left(\sqrt{2}\pi\right)^{1/3}|q_{I}Y^{I}| \, .
\end{eqaed}
A similar expression holds for magnetic BPS string tensions. The~fact that no such state becomes light at large moduli reflects their M-theoretic origin in terms of M2-branes wrapped on two-cycles. Thus, we can anticipate that no infinite-distance limit of this type will badly break the one-form symmetry $U(1)^{h^{1,1}(X)}_\text{el}$, and~one would be left with the more complicated and model-dependent limits discussed in, e.g.,~\cite{Alim:2021vhs}. The~most promising candidates for a general analysis along the lines of the preceding section are perhaps particles with charges $q_{0}$ associated with the graviphoton $A^{0}$ since, when moving in the vector multiplet moduli space, their masses are
\begin{eqaed}
 m(q_{0})=\left(\sqrt{2}\pi\right)^{1/3}q_{0} \, .
\end{eqaed}
According to our general discussion in \cref{sec:bb_dd}, the~bad-breaking parameter would be given by
\begin{eqaed}
    \delta_\Lambda = \sum_{\{q_{0}\}}\frac{(g^{2}\Lambda)q_{0}^{4}}{4\pi^{2}\Lambda^{2}}\int_{0}^{\infty}ds W_{1}(s)\left(\left(\frac{sq_{0}}{\Lambda}\right)^{2}J_{3}\left( \frac{sq_{0}}{\Lambda}\right)+ \left(\frac{4sq_{0}}{\Lambda}\right)J_{2}\left(\frac{sq_{0}}{\Lambda}\right)\right).
\end{eqaed}
If the above infinite sum converges, the~overall scaling at the Planck scale would appear to be that of $g^2$, meaning that, once again, bad breaking would not occur in this toy model unless $C_{00K}=0$ for $K>0$. The~same conclusion holds if the sum diverges: in this case, BPS masses are to be cut off at the Planck scale, which is the species scale in the limits of interest, since no light species appear. This means $q_0 = \mathcal{O}(1)$, leading to the same overall scaling with the gauge coupling. If~$C_{000}$ is the only non-vanishing intersection number of this type, the~one-form symmetry associated with the graviphoton is badly broken at the Planck~scale.

\section{Conclusions}\label{sec:conclusions}

In this work, we developed and computed a refined version of the bad-breaking parameter defined in~\cite{Cordova:2022rer} that is applicable in general contexts, including mixed Abelian gauge groups in any dimensions. This framework allows one to quantify the well-known and kinematically substantiated notion of absence of global symmetries in Quantum Gravity. Specifically, we examined how towers of BPS states break the one-form symmetries in four-dimensional and five-dimensional supergravity. As~a proof of concept of the validity of this framework in this well-studied, top-down setting, we performed some computations in toy models with a single modulus and constant BPS degeneracies, laying the groundwork for a more systematic numerical study. While most of these toy models do not seem to display the expected bad breaking condition, this may be an artifact of the simplifying assumptions we made along the way. Indeed, one of the main lessons of the swampland program is precisely that consistent, top-down constructions have several highly non-generic features. This calls for further investigations of bad breaking in the string landscape. In~order to do so, it would be necessary to study settings with additional moduli and/or non-simplicial cones of charges. To~achieve this, specialized software such as \texttt{CYTools 1.2.6} would provide an invaluable asset. In~particular, a~more accurate treatment of BPS degeneracies is necessary, starting by estimating them via Gopakumar--Vafa~invariants.

In the same setting of Calabi--Yau compactifications, it would be interesting to explore the bad breaking condition for large numbers of moduli or large intersection numbers. The~existence of these limits would contradict the alleged finiteness of inequivalent Calabi--Yau threefolds, which would be implied by the finiteness of the string landscape, as recently investigated via swampland considerations in~\cite{Kim:2024hxe}. Finding universal physical inconsistencies in such putative limits would provide another physical rationale for~finiteness.

Beyond a direct extension of our analysis, there are several additional research avenues to pursue following this work. Firstly, it would be fascinating to adapt or expand this framework to investigate settings in anti-de Sitter spacetimes, focusing on the implications for dual conformal field theories. The~bad breaking condition may be able to be recast in these terms, potentially allowing for sharp statements along the lines of those in~\cite{Harlow:2018jwu, Harlow:2018tng}. 

Another option is to explore the role of the weak-gravity scale $\Lambda_{U(1)}$ as a possible cutoff, in~the spirit of~\cite{Heidenreich:2017sim}, and~whether the relevant condition for symmetry breaking ought to be modified. Specifically, according to~\cite{Heidenreich:2017sim}, in~this approach, one would exclude contributions from particles with masses with momenta and masses in the range $p \lesssim m \lesssim \Lambda_{U(1)}$. This exclusion is justified in~\cite{Heidenreich:2017sim} by the fact that the impact of heavy particles diminishes as $p$ approaches the scale $\Lambda_{U(1)}$, where $|\Pi(p^{2})|$ is approximately one. Moreover, there are relatively few particles within this narrow range. Their influence can be approximated by modifying $\Pi(p^{2})\to \Pi(p^{2})+ p^{2}/\Lambda_{U(1)}^{2}$. We look forward to further pursuing this approach in future~work.

In this study we did not address higher-form symmetries. For~instance, magnetic two-form symmetries in five dimensions could be examined. Similarly, we did not discuss non-invertible symmetries~\cite{Schafer-Nameki:2023jdn, Shao:2023gho, Costa:2024wks}. It would be intriguing to explore the non-invertible symmetries present in 11d and 10d type II supergravities~\cite{Heidenreich:2020pkc, Fernandez-Melgarejo:2024ffg, Garcia-Valdecasas:2023mis} through the lens of a bad-breaking scenario. At~the kinematic level, it is known that these symmetries are broken by gravitational solitons~\cite{McNamara:2021cuo, Heckman:2024obe} and,~thus, the dynamical expectation is for bad breaking to occur at the Planck scale. Generalized symmetries are also intimately connected with axions and generalized theta terms~\cite{McNamara:2020uza, Grimm:2022xmj}, which may open up an enticing window into phenomenology~\cite{Choi:2023pdp, Dierigl:2024cxm}.

Lastly, a~fascinating albeit somewhat ambiguous question is whether alternative methods for quantifying the breaking of symmetries may capture different aspects of the physics and~perhaps exhibit different properties. If~a systematic understanding of symmetry-breaking quantification would lead to a universal bad-breaking condition, the~emerging picture would shed light on the role of global symmetries---or lack thereof---in Quantum~Gravity.

\vspace{6pt} 

%%%%%%%%%%%%%%%%%%%%%%%%%%%%%%%%%%%%%%%%%%
%% optional
%\supplementary{The following supporting information can be downloaded at:  \linksupplementary{s1}, Figure S1: title; Table S1: title; Video S1: title.}


% Only used for preprtints:
% \supplementary{The following supporting information can be downloaded at the website of this paper posted on \href{https://www.preprints.org/}{Preprints.org}.}

%\begin{tabularx}{\textwidth}{lll}
%\toprule
%\textbf{Name} & \textbf{Type} & \textbf{Description} \\
%\midrule
%S1 & Python script (.py) & Script of python source code used in XX \\
%S2 & Text (.txt) & Script of modelling code used to make Figure X \\
%S3 & Text (.txt) & Raw data from experiment X \\
%S4 & Video (.mp4) & Video demonstrating the hardware in use \\
%\bottomrule
%\end{tabularx}
%}

%%%%%%%%%%%%%%%%%%%%%%%%%%%%%%%%%%%%%%%%%%
\authorcontributions{{Conceptualization, I.B. and P.G.; methodology, I.B. and P.G.; software, I.B. and P.G.; validation, I.B. and P.G.; formal analysis, I.B. and P.G.; investigation, I.B. and P.G.; resources, I.B. and P.G.; writing---original draft preparation, I.B. and P.G.; writing---review and editing, I.B. and P.G.; visualization, I.B. and P.G.; supervision, I.B.; project administration, I.B. and P.G.; funding acquisition, I.B. All authors have read and agreed to the published version of the manuscript.}}

\funding{The work of I.B. is supported by the Origins Excellence Cluster and the German-Israel-Project (DIP) on Holography and the Swampland.}

%\institutionalreview{}
%
%\informedconsent{}

\dataavailability{{No new data were created or analyzed in this study.}}

\acknowledgments{The authors are grateful to N. Gendler and A. Gnecchi for discussions and collaboration on a related~project.}

\conflictsofinterest{{The authors declare no conflict of interest.}} 

%%%%%%%%%%%%%%%%%%%%%%%%%%%%%%%%%%%%%%%%%%
%% Optional

%% Only for journal Encyclopedia
%\entrylink{The Link to this entry published on the encyclopedia platform.}

%%%%%%%%%%%%%%%%%%%%%%%%%%%%%%%%%%%%%%%%%%
%% Optional
\appendixtitles{yes} % Leave argument "no" if all appendix headings stay EMPTY (then no dot is printed after "Appendix A"). If~the appendix sections contain a heading then change the argument to "yes".
\appendixstart
\appendix
\section[\appendixname~\thesection]{\textbf{Supergravity in Five Dimensions and BPS States}}\label{app:A}

In this appendix, we briefly review some background material on $5d$ supergravity, of~which we make extensive use in the main text. Compactifying M-theory on a Calabi--Yau threefold $X$, the~field content of the resulting five-dimensional $\mathcal{N}=1$ supergravity comprises the gravity multiplet, $n=h^{1,1}(X)-1$ vector multiplets, and~$h^{2,1}(X)+ 1$ hypermultiplets. The~K\"{a}hler moduli of $X$ are $n$ scalars characterizing the vector multiplet moduli space. As~such, in~a generic locus in moduli space, the gauge group is $U(1)^{n+1}$, where the extra $U(1)$ arises from the graviphoton. Choosing a basis of divisors ${[H_{a}]}_{a=1}^{h^{1,1}(X)}$, one can write down the K\"{a}hler form of $X$ in terms of projective coordinates of the vector multiplet moduli space according to
\begin{equation}
    \mathcal{J}= \sum_{a=1}^{h^{1,1}(X)}t^{a}[H_{a}] \, .
\end{equation}
The bosonic part of the effective action, neglecting hypermultiplets, is given by
\begin{adjustwidth}{-\extralength}{0cm}
\begin{equation}\label{eq:5d_action}
    S= \frac{1}{2\kappa_{5}^{2}}\int d^{5}x\sqrt{-g}\left(R- \frac{1}{2}g_{ij}(\phi)\partial \phi^{i}\cdot \partial \phi^{j}\right)-\frac{1}{2g_{5}^{2}}\int a_{IJ}(\phi)F^{I}\wedge \star F^{J}+ \frac{1}{24\pi^{2}}\int C_{IJK}A^{I}\wedge F^{J}\wedge F^{K} \, .
\end{equation}
\end{adjustwidth}
The prepotential $\mathcal{F}(Y)$ encodes the gauge kinetic matrix $a_{IJ}(\phi)$, the~moduli--space metric $g_{ij}(\phi)$, and the Chern--Simons couplings $C_{IJK}$. Since the overall internal volume is part of the universal hypermultiplet, the~vector-multiplet moduli space is defined by the constraint
\begin{equation}
    \mathcal{F}(Y(\phi))=1 \, ,
\end{equation}
which can be parametrized by local coordinates $\phi^a$ (vacuum expectation values of the scalar fields in \cref{eq:5d_action}). The~exact prepotential can be expressed as a cubic homogeneous polynomial in terms of the triple intersection numbers $C_{IJK}=\int_{X}[H_{I}]\wedge [H_{J}]\wedge [H_{K}]$ according to
\begin{equation}
    \mathcal{F}=\frac{1}{6}C_{IJK}Y^{I}Y^{J}Y^{K} \, ,
\end{equation}
from which
\begin{equation}
    a_{IJ}(\phi)= \mathcal{F}_{I}[Y(\phi)]\mathcal{F}_{J}[Y(\phi)]- \mathcal{F}_{IJ}[Y(\phi)], \quad g_{ij}(\phi)=a_{IJ}(\phi)\partial_{i}Y^{I}\partial_{j}Y^{J},
\end{equation}
where we define $\mathcal{F}_{I}=\partial_{I}\mathcal{F}$ and $\mathcal{F}_{IJ}=\partial_{I}\partial_{J}\mathcal{F}$. Moreover, the~metric of the moduli space can be recast as
\begin{equation}
    g_{IJ}= \frac{2}{3}\frac{\mathcal{F}_{I}\mathcal{F}_{J}}{\mathcal{F}^{2}}- \frac{\mathcal{F}_{IJ}}{\mathcal{F}}\, .
\end{equation}

Coupling this theory with a massive charged particle described by a probe action
\begin{equation}
    S_{pp}= -\int m(\phi)d\tau + q_{I}\int A^{I} \, ,
\end{equation}
its charged $q_{I}$ is determined by Gauss' law via the flux integral:
\begin{equation}
    q_{I}=\oint_{S^{3}}\left(\frac{1}{g_{5}^{2}}a_{IJ}(\phi)\star F^{J}-\frac{1}{2(2\pi)^{2}}C_{IJK}A^{J}\wedge F^{K} \right) .
\end{equation}
Such particles satisfy the BPS bound
\begin{equation}
    m(q_{I})\geq \left(\frac{\sqrt{2}\pi}{\kappa_{5}}\right)^{1/3}|Z|= \left(\frac{\sqrt{2}\pi}{\kappa_{5}}\right)^{1/3}\frac{|q_{I}Y^{I}|}{\mathcal{F}^{1/3}} \, , \qquad |Z|\equiv\frac{|q_{I}Y^{I}|}{\mathcal{F}^{1/3}} \, ,
\end{equation}
given by the central charge of the supersymmetry algebra. In~the main text, we use Planck units in which $\kappa_5 = 1$.

\section[\appendixname~\thesection]{\textbf{Supergravity in Four Dimensions}}\label{app:B}

In this appendix, we provide a brief overview of some basic features of the four-dimensional $\mathcal{N}=2$ supergravity employed in the main text. Realizing these theories compactifying the type IIA string on a Calabi--Yau threefold $X$, K\"{a}hler structure deformations $t^{I}$ parametrize the moduli space of vector multiplets. These moduli can be defined by expanding the complexified K\"{a}hler form in a cohomological basis:
\begin{equation}
    B_{2}+i\mathcal{J}= t^{I}\omega_{I}, \quad I=1,\dots, h^{1,1}(X).
\end{equation}
Here, $B_{2}$ expressed as $B_{2}=b^{I}\omega_{I}$ is the NS-NS two-form. Writing $\mathcal{J} = v^{I}\omega_{I}$, we have $t^{I}= b^{I}+ iv^{I}$. Similar to the five-dimensional case, the~triple intersection numbers
\begin{equation}
   C_{IJK}=\int_{Y_{3}}\omega_{I}\wedge \omega_{J}\wedge \omega_{K}
\end{equation}
appear in the volume (in string units) and classical K\"{a}hler potential
\begin{equation}
    \mathcal{V}= \frac{1}{3!}\int_{Y_{3}}J\wedge J\wedge J= \frac{1}{6}C_{IJK}v^{I}v^{J}v^{K} \, , \qquad K=-\log 8\mathcal{V} \, .
\end{equation}

In the main text, we consider type IIA string theory compactified on Calabi--Yau threefold $X$. The~massless field content includes symplectic vectors $A=(\Tilde{A}^{I}, A_{I})$ and elements of the even cohomology $H^{2\star}(X)= H^{0}(X)\oplus H^{(1,1)(X)}\oplus H^{(2,2)(X)} \oplus H^{(6)(X)}$, which have $(2h^{1,1}(X)+2)$-components in total. One can expand these vectors on a harmonic basis $(\alpha_{I}, \beta^{I})=(\alpha_{0}, \alpha_{A}, \beta^{A}, \beta^{0})$ of even cohomology:
\begin{equation}
    A=(\Tilde{A}^{I}, A_{I})= \Tilde{A}^{I}\alpha_{I}+ A_{I}\beta^{I} \, ,
\end{equation}
where the basis can be chosen to be symplectic with respect to the intersection product:
\begin{equation}
    \langle \alpha_{I}, \beta^{J}\rangle= \int_{X} \alpha_{I}\wedge \beta^{J\star}=\delta_{I}^{J}\, .
\end{equation}
The dual homological description of these vectors provides a physical interpretation of the BPS states charged under them. Namely, they comprise (bound states of) D0-branes located at a point in the Calabi--Yau, D2-branes wrapped around two-cycles, D4-branes wrapped around four-cycles, and D6-branes wrapped around the full~Calabi--Yau. 

%%%%%%%%%%%%%%%%%%%%%%%%%%%%%%%%%%%%%%%%%%
%\isPreprints{}{% This command is only used for ``preprints''.
\begin{adjustwidth}{-\extralength}{0cm}

\reftitle{References}

\PublishersNote{}
%\isPreprints{}{% This command is only used for ``preprints''.
\end{adjustwidth}
\end{document}